\documentclass[english,10pt,letterpaper]{article}
\usepackage[T1]{fontenc}
\usepackage[left=35mm, right=35mm, top=30mm, bottom=30mm]{geometry}
\usepackage{graphicx}
\usepackage{authblk}
\usepackage{amssymb}
\usepackage{amsmath}
\usepackage{babel}
\usepackage{indentfirst}
\usepackage{setspace}
\usepackage{lettrine}
\usepackage{fancyhdr}
\usepackage[labelfont=bf]{caption}
\providecommand{\indexterms}[1]{\small{\textbf{\textit{Index Terms---}}}#1}
\providecommand{\abst}[1]{\textbf{\textit{Abstract---}}#1}

\title{\Huge{\textbf{Noise Analysis for Performance Evaluation of Biopotential Recording Front-Ends}}}

\author{Taeju Lee\thanks{T. Lee was with the School of Electrical Engineering, Korea Advanced Institute of Science and Technology (KAIST), Daejeon 34141, Republic of Korea. He is now with the Department of Electrical Engineering, Columbia University, New York, NY 10027, USA.}}

\affil{\small{Department of Electrical Engineering, Columbia University, New York, NY 10027, USA\\
	E-mail: taeju.lee@columbia.edu}}
	
\date{}

\begin{document}
\maketitle
\pagestyle{fancy}
\fancyhf{}
\lhead{\small{LEE: NOISE ANALYSIS FOR PERFORMANCE EVALUATION OF BRFE}}
\cfoot{\thepage}

\noindent
\small{\abst{\textbf{Noise efficiency factor and power efficiency factor are widely used as the figure of merit to quantify the performance of biopotential recording front-ends. The noise efficiency factor provides a dimensionless quantity that indicates how much better current-noise efficiency is achieved compared to an ideal bipolar junction transistor when assuming the total bias current and recording bandwidth of a front-end are the same as an ideal bipolar junction transistor. The power efficiency factor indicates how much better power-noise efficiency is achieved compared to an ideal bipolar junction transistor, taking into account the supply voltage, total bias current, and bandwidth. Compared to the noise efficiency factor, the power efficiency factor is calculated using voltage and current values, thereby enabling the power consumption $(= \textit{V\textsubscript{DD}}\times\textit{I\textsubscript{DD}})$-considered noise performance comparison between different front-ends. In this article, the noise efficiency factor and power efficiency factor of biopotential recording front-ends are explored according to CMOS technology scaling and front-end architectures.}}}
\singlespacing
\noindent
\small{\indexterms{\textbf{Analog front-end, biopotential amplifier, bioelectric signal, brain-machine interface, BMI, brain-computer interface, BCI, CMOS, figure of merit, FoM, noise analysis, noise efficiency factor, NEF, power efficiency factor, PEF, technology scaling.}}}

\singlespacing
\singlespacing
\singlespacing

\section{\Large{I}\large{NTRODUCTION}}
\lettrine[findent=2pt, nindent=0pt]{\textbf{B}}{\textbf{iopotential}} recording front-ends have played an important role in observing bioelectric signals, which has revolutionized biomedical engineering spanning personal healthcare, neuroscience, brain-machine interface, etc. Over the last decades, the performance of biopotential recording front-ends has considerably advanced in the input-referred noise, design area, energy efficiency, etc. In reality, the design area of a single front-end has been reduced, the power consumption has also been reduced as the supply voltage has declined, and nonlinearity has greatly improved. In addition, various front-end performances have significantly advanced, enabling reliable \textit{in-vivo} and \textit{in-vitro} signal monitoring.

\indent
The performance of biopotential recording front-ends has been quantified using the common-mode rejection ratio (CMRR), power supply rejection ratio (PSRR), dynamic range (DR), input impedance, etc. Among them, the noise efficiency factor (NEF) and power efficiency factor (PEF) have been widely employed to quantify the performance of front-ends over the last decades.\\

\begin{figure}[h]
	\centering
	\includegraphics[scale=0.5]{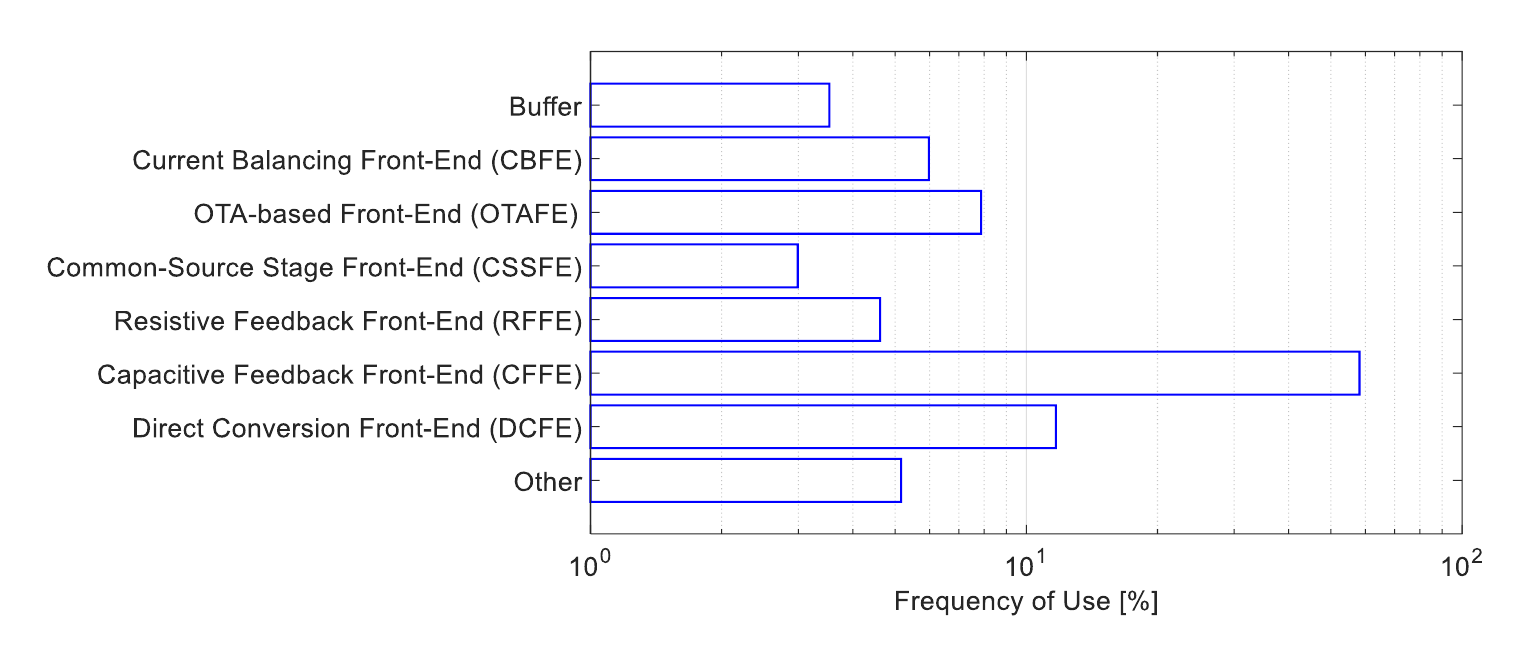}
	\caption{Frequency of use of each front-end architecture.}
	\label{Fig. 1}
\end{figure}

\indent
NEF and PEF have been widely used as figure of merits (FoMs) to summarize and compare the front-end performance by taking into account the input-referred noise, bandwidth, bias current, and supply voltage. Since the 1970s, various kinds of front-end architectures have been developed to observe bioelectric signals and appropriate architecture has been used to meet the target performance. Depending on the front-end architecture, NEF and PEF have been diversely achieved. Fig. \ref{Fig. 1} presents how frequently each front-end architecture has been used over the past decades. Note that Fig. \ref{Fig. 1} is implemented by investigating the architectures of biopotential recording front-ends developed since the 1970s and the front-ends investigated are listed in \cite{r1}, \cite{r2}.

\indent
As shown in Fig. \ref{Fig. 1}, a variety of architectures have been used to design the front-ends, but among those architectures, the capacitive feedback front-end \cite{r3} has been widely used due to its DC offset filtering, superior noise performance, and design simplicity. Also, diverse circuit techniques have been developed based on the capacitive feedback architecture to improve NEF and PEF. Moreover, to overcome the theoretical limit of NEF and PEF in the capacitive feedback front-end, the front-ends based on discrete-time amplification \cite{r4}, \cite{r5} are proposed, which is classified as \emph{Other} in Fig. \ref{Fig. 1} since the number of prototypes is not yet sufficient.

\indent
In this article, NEF and PEF of biopotential recording front-ends developed since the 1970s are explored according to the architectures categorized in Fig. \ref{Fig. 1}. NEF and PEF used in this article are obtained by investigating the prior front-ends listed in \cite{r1}, \cite{r2}. Section II describes the NEF background as well as the trends according to the architectures and technology nodes. The background and trends of PEF are presented in Section III. Section IV presents the performance mapping of technology nodes, NEF, and PEF for all front-end architectures. Finally, Section V concludes this article by discussing other FoMs that need to be considered in front-end design. Note that NEF and PEF are used to quantify the performance of a front-end that measures voltage signals rather than current signals, therefore, the front-ends considered in this article mean tools measuring voltage signals generated from brain activity, heart activity, muscle activity, etc.\\

\section{\Large{N}\large{OISE} \Large{E}\large{FFICIENCY} \Large{F}\large{ACTOR}}
The biopotential recording front-end, which measures voltage signals, has the input-referred noise $\textit{V\textsubscript{n,rms}}$. $\textit{V\textsubscript{n,rms}}$ is the integrated noise voltage over the recording bandwidth $\textit{BW\textsubscript{fe}}$ while consuming the total bias current $\textit{I\textsubscript{DD}}$ and is shown at the input port of a front-end. To quantify and compare the performance of different front-ends, the NEF is calculated using $\textit{V\textsubscript{n,rms}}$, $\textit{BW\textsubscript{fe}}$, and $\textit{I\textsubscript{DD}}$. The NEF is defined by comparing the input-referred noise voltage of a front-end to that of a bipolar junction transistor (BJT), assuming that the bias current and bandwidth of a BJT are equal to $\textit{I\textsubscript{DD}}$ and $\textit{BW\textsubscript{fe}}$, respectively, as described in \cite{r6}. The NEF proposed in \cite{r6} is expressed as $\textit{V\textsubscript{n,rms}}{\sqrt{2\textit{I\textsubscript{DD}}/(\pi\cdot\textit{U\textsubscript{T}}\cdot4kT\cdot{\textit{BW\textsubscript{fe}}})}}$, where $\textit{U\textsubscript{T}}$ is the thermal voltage $kT/q$, $k$ is the Boltzmann constant, and $T$ is the absolute temperature.

\subsection{\textit{BJT Noise Calculation}}
In this article, when comparing the noise voltage of a front-end to that of a BJT, the calculation is conducted using either a single-ended common-emitter stage or a differential common-emitter stage. The input-referred noise voltage of a BJT is calculated as follows. Assuming that the noise voltage of a BJT is considered only using the collector shot noise $i\rlap{\textsuperscript{2}}\textit{\textsubscript{c}}$ while neglecting the base shot noise, base resistance, and flicker noise, the input noise voltage of a single-ended common-emitter stage is expressed as $i\rlap{\textsuperscript{2}}\textit{\textsubscript{c}}/g\rlap{\textsuperscript{2}}\textit{\textsubscript{m}}$, where $\textit{g\textsubscript{m}}$ is the BJT transconductance expressed as $\textit{I\textsubscript{C}}/\textit{U\textsubscript{T}} = q\textit{I\textsubscript{C}}/kT$. Note that $i\rlap{\textsuperscript{2}}\textit{\textsubscript{c}} = 2q\textit{I\textsubscript{C}}$, where $q$ and $\textit{I\textsubscript{C}}$ are the electron charge and collector bias current, respectively. Therefore, the input-referred noise voltage of a single-ended common-emitter stage $V\rlap{\textsuperscript{2}}\textit{\textsubscript{n,single-bjt}}$ can be derived as $i\rlap{\textsuperscript{2}}\textit{\textsubscript{c}}/g\rlap{\textsuperscript{2}}\textit{\textsubscript{m}} = 2q\textit{I\textsubscript{C}}/g\rlap{\textsuperscript{2}}\textit{\textsubscript{m}} = 2\textit{g\textsubscript{m}}kT/g\rlap{\textsuperscript{2}}\textit{\textsubscript{m}} = 2kT/\textit{g\textsubscript{m}}$ and lead to

\begin{align}
	V\rlap{\textsuperscript{2}}\textit{\textsubscript{n,single-bjt}} &= \frac{2kT}{\textit{g\textsubscript{m}}}\\
	&= \frac{2kT\cdot\textit{U\textsubscript{T}}}{\textit{I\textsubscript{C}}}
\end{align}

\noindent
Since the noise voltage squared of a differential pair is twice that of a single-ended stage as described in \cite{r7}, the input-referred noise voltage of a differential common-emitter stage $V\rlap{\textsuperscript{2}}\textit{\textsubscript{n,diff-bjt}}$ is twice $V\rlap{\textsuperscript{2}}\textit{\textsubscript{n,single-bjt}}$ and is written as

\begin{align}
	V\rlap{\textsuperscript{2}}\textit{\textsubscript{n,diff-bjt}} &= \frac{4kT}{\textit{g\textsubscript{m}}}\\
	&= \frac{4kT\cdot\textit{U\textsubscript{T}}}{\textit{I\textsubscript{C}}}
\end{align}

\indent
Considering that the noise bandwidth can be approximated as $\pi/2$ times the recording bandwidth in a one-pole recording front-end as described in \cite{r7}, the total input-referred noise voltage of each stage over the recording bandwidth can be expressed as

\begin{align}
	V\rlap{\textsuperscript{2}}\textit{\textsubscript{n,single-bjt,tot}} &= \int_{0}^{\infty}V\rlap{\textsuperscript{2}}\textit{\textsubscript{n,single-bjt}}\,df\\
	&= \frac{2kT\cdot\textit{U\textsubscript{T}}}{\textit{I\textsubscript{C}}}\cdot{\textit{BW\textsubscript{bjt}}}\cdot\frac{\pi}{2} 
\end{align}

\begin{align}
	V\rlap{\textsuperscript{2}}\textit{\textsubscript{n,diff-bjt,tot}} &= \int_{0}^{\infty}V\rlap{\textsuperscript{2}}\textit{\textsubscript{n,diff-bjt}}\,df\\
	&= \frac{4kT\cdot\textit{U\textsubscript{T}}}{\textit{I\textsubscript{C}}}\cdot{\textit{BW\textsubscript{bjt}}}\cdot\frac{\pi}{2}
\end{align}

\noindent
where $\textit{BW\textsubscript{bjt}}$ is the noise bandwidth of the BJT. The total input-referred noise voltage of each BJT stage can be rewritten as

\begin{align}
	V\textit{\textsubscript{n,single-bjt,tot}} = \sqrt{\frac{2kT\cdot\textit{U\textsubscript{T}}}{\textit{I\textsubscript{C}}}\cdot{\textit{BW\textsubscript{bjt}}}\cdot\frac{\pi}{2}}
	\,\,\,
	\text{and}
	\,\,\,
	V\textit{\textsubscript{n,diff-bjt,tot}} = \sqrt{\frac{4kT\cdot\textit{U\textsubscript{T}}}{\textit{I\textsubscript{C}}}\cdot{\textit{BW\textsubscript{bjt}}}\cdot\frac{\pi}{2}} 
\end{align}

\indent
Note that a front-end achieves $\textit{V\textsubscript{n,rms}}$ over $\textit{BW\textsubscript{fe}}$ while consuming $\textit{I\textsubscript{DD}}$. Assuming $\textit{BW\textsubscript{fe}} = \textit{BW\textsubscript{bjt}}$ and $\textit{I\textsubscript{DD}} = \textit{I\textsubscript{C}}$ (in other words, the bandwidth and bias current are equalized across the front-end and BJT to compare the noise performance fairly while applying the same design parameters), the noise performance of a fabricated front-end can be compared to the noise voltage of the BJT as follows:

\begin{align}
	\frac{V\textit{\textsubscript{n,rms}}}{V\textit{\textsubscript{n,single-bjt,tot}}} = V\textit{\textsubscript{n,rms}}\sqrt{\frac{\textit{I\textsubscript{DD}}}{\pi\cdot\textit{U\textsubscript{T}}\cdot{kT}\cdot{\textit{BW\textsubscript{fe}}}}}
	\,\,
	\text{and}
	\,\,
	\frac{V\textit{\textsubscript{n,rms}}}{V\textit{\textsubscript{n,diff-bjt,tot}}} = V\textit{\textsubscript{n,rms}}\sqrt{\frac{\textit{I\textsubscript{DD}}}{\pi\cdot\textit{U\textsubscript{T}}\cdot2kT\cdot{\textit{BW\textsubscript{fe}}}}}
\end{align}
 
\indent
In the calculation of $\textit{V\textsubscript{n,single-bjt,tot}}$ and $\textit{V\textsubscript{n,diff-bjt,tot}}$, the collector shot noise is only considered while ignoring the base shot noise, base resistance, and flicker noise. However, in reality, $\textit{V\textsubscript{n,rms}}$ is measured while including the thermal and flicker noise. $\textit{V\textsubscript{n,rms}}/\textit{V\textsubscript{n,single-bjt,tot}}$ and $\textit{V\textsubscript{n,rms}}/\textit{V\textsubscript{n,diff-bjt,tot}}$ mean how much better the noise performance of a front-end is achieved when compared to a single BJT and a differential BJT, respectively, in the equalized bandwidth and bias current. The concept of calculating $\textit{V\textsubscript{n,rms}}/\textit{V\textsubscript{n,single-bjt,tot}}$ and $\textit{V\textsubscript{n,rms}}/\textit{V\textsubscript{n,diff-bjt,tot}}$ is the same as the NEF definition in \cite{r6}. However, the noise calculation is conducted according to the single and differential types in this article. Defining $\textit{NEF\textsubscript{single-bjt}} = \frac{\textit{V\textsubscript{n,rms}}}{\textit{V\textsubscript{n,single-bjt,tot}}}$ and $\textit{NEF\textsubscript{diff-bjt}} = \frac{\textit{V\textsubscript{n,rms}}}{\textit{V\textsubscript{n,diff-bjt,tot}}}$, $\textit{NEF\textsubscript{single-bjt}} = \sqrt{2}\textit{NEF\textsubscript{diff-bjt}}$. Also, compared to the NEF definition of $\textit{V\textsubscript{n,rms}}{\sqrt{2\textit{I\textsubscript{DD}}/(\pi\cdot\textit{U\textsubscript{T}}\cdot4kT\cdot{\textit{BW\textsubscript{fe}}})}}$, the results from Eq. (10) can be also expressed as $\textit{NEF\textsubscript{single-bjt}} = \sqrt{2}\text{NEF}$ and $\textit{NEF\textsubscript{diff-bjt}} = \text{NEF}$.\\

\subsection{\textit{NEF Trends}}
\begin{figure}[h]
	\centering
	\includegraphics[scale=0.55]{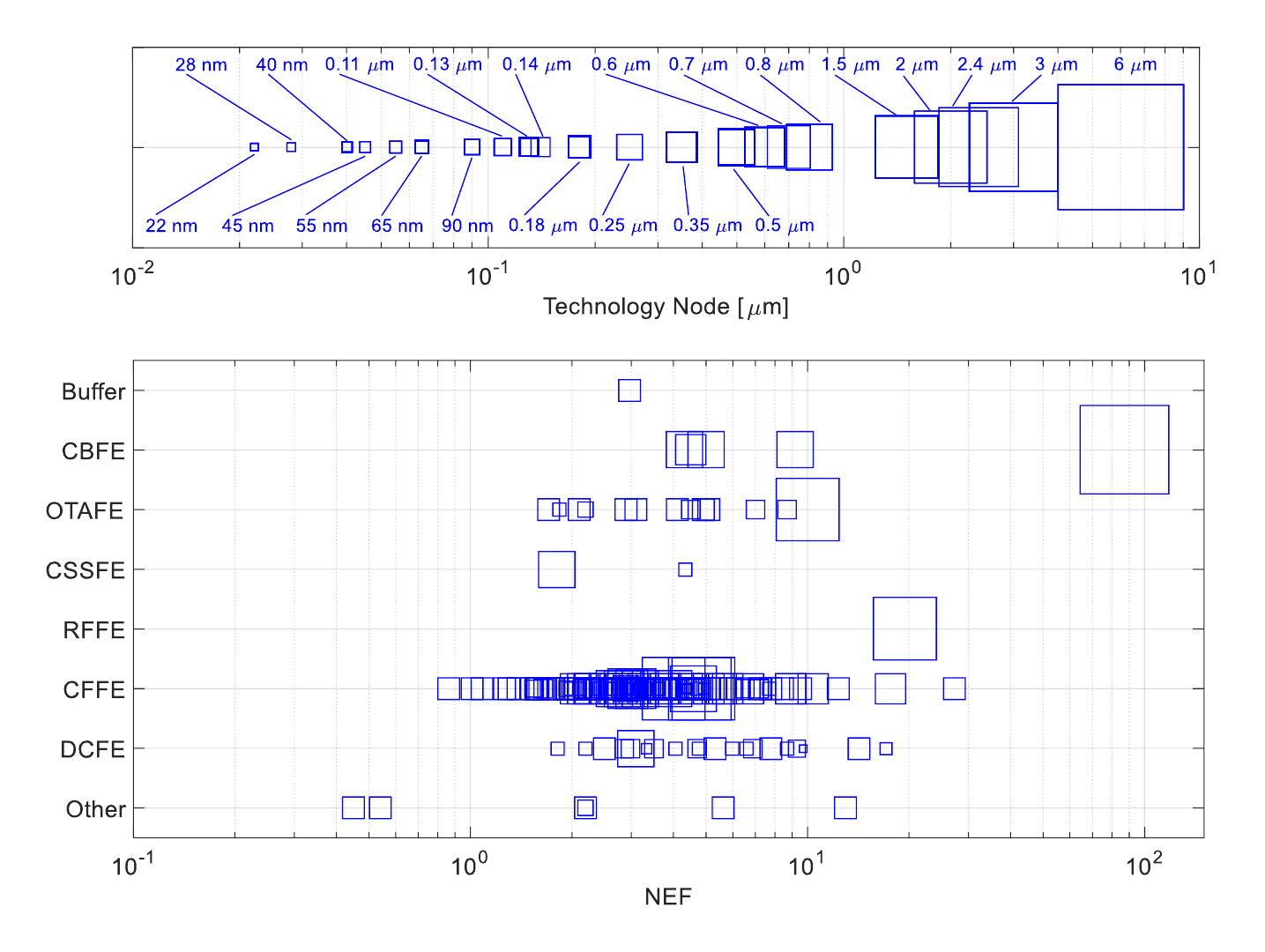}
	\caption{NEF according to front-end architectures and technology nodes.}
	\label{Fig. 2}
\end{figure}

Over the last decades, biopotential recording front-ends have been developed using various circuit techniques, architectures, and technology nodes for improving noise performance while consuming low power. These efforts have led to advances in the NEF. Fig. \ref{Fig. 2} shows the overall NEF trends according to the front-end architectures. Each NEF is expressed by using a square symbol and the symbol size indicates a technology node used. The top of Fig. \ref{Fig. 2} summarizes the symbol sizes representing the technology nodes ranging from 6 $\mu$m to 22 nm. As complementary metal-oxide semiconductor (CMOS) technology nodes have scaled down, the performance of digital circuits has also advanced by benefiting from scaling. However, most NEFs have not advanced as much as the digital circuits have benefited from scaling, as discussed in \cite{r1} and shown in Fig. \ref{Fig. 2}. To improve the NEF, a front-end needs to achieve a low $\textit{V\textsubscript{n,rms}}$ while consuming a low $\textit{I\textsubscript{DD}}$ given $\textit{BW\textsubscript{fe}}$. However, in most CMOS-based biopotential recording front-ends, the thermal and flicker noise trades with the bias current and transistor area. Therefore, the NEF improvement is a challenging task that requires optimized design techniques.

\indent
As shown in the bottom of Fig. \ref{Fig. 2}, diverse front-end architectures have achieved different NEFs. In particular, as discussed in Fig. \ref{Fig. 1}, the capacitive feedback front-end (CFFE) is overwhelmingly employed among the architectures and various circuit techniques for the CFFE have been developed to reduce the NEF. Also, to overcome the theoretical limit of NEF in the CFFE, discrete-time amplification-based front-ends \cite{r4}, \cite{r5} are developed. However, since the number of discrete-time amplification-based front-ends is not yet sufficient, it is classified as \emph{Other} in Fig. \ref{Fig. 2}. Among numerous front-ends, the CFFEs have mostly achieved the NEF between 1 to 10 using a variety of technology nodes. Compared to the front-ends which require an independent analog-to-digital converter (ADC), a direct conversion front-end (DCFE) digitizes the input signal directly while incorporating the function of analog-to-digital conversion into its architecture. The DCFE is the second most used architecture after the CFFE, as shown in Fig. \ref{Fig. 1}, and achieves the NEF between 1 to 10, mostly using sub-0.35 $\mu$m nodes.

\indent
Although numerous biopotential recording front-ends have been developed since the 1970s, the concept of NEF \cite{r6} was presented in the 1980s and there are also many literatures that do not specify the NEF. Therefore, NEFs noted in Fig. \ref{Fig. 2} do not reflect all front-ends shown in Fig. \ref{Fig. 1}.\\

\section{\Large{P}\large{OWER} \Large{E}\large{FFICIENCY} \Large{F}\large{ACTOR}}
\subsection{\textit{Motivation}}
Revisiting the definition of NEF, $\textit{V\textsubscript{n,rms}}{\sqrt{2\textit{I\textsubscript{DD}}/(\pi\cdot\textit{U\textsubscript{T}}\cdot4kT\cdot{\textit{BW\textsubscript{fe}}})}}$, the only parameter related to power consumption is $\textit{I\textsubscript{DD}}$ while neglecting the supply voltage $\textit{V\textsubscript{DD}}$. Therefore, the NEF can provide a fair performance comparison between different front-ends when the front-ends being compared employ the same $\textit{V\textsubscript{DD}}$. In other words, although the NEF can compare fairly the current-noise efficiency between the front-ends, the power-noise efficiency between different front-ends cannot be provided fairly using the NEF if the front-ends are developed using different supply voltages. To overcome this limit, the PEF was developed in \cite{r8}. The PEF is calculated by squaring the NEF and multiplying it by $\textit{V\textsubscript{DD}}$, meaning $\text{PEF} = \text{NEF}\textsuperscript{2}\textit{V\textsubscript{DD}} = V\rlap{\textsuperscript{2}}\textit{\textsubscript{n,rms}}\frac{2\textit{I\textsubscript{DD}}\textit{V\textsubscript{DD}}}{\pi\cdot\textit{U\textsubscript{T}}\cdot4kT\cdot\textit{BW\textsubscript{fe}}} = V\rlap{\textsuperscript{2}}\textit{\textsubscript{n,rms}}\frac{2\textit{P\textsubscript{tot}}}{\pi\cdot\textit{U\textsubscript{T}}\cdot4kT\cdot\textit{BW\textsubscript{fe}}}$, where $\textit{P\textsubscript{tot}}$ is the total power consumption of a front-end. The PEF contains $\textit{P\textsubscript{tot}} = \textit{I\textsubscript{DD}}\textit{V\textsubscript{DD}}$, allowing us to compare the power-noise efficiency between different front-ends.

\indent
Rethinking the definition of PEF as discussed in \cite{r8}, the PEF is defined by comparing $V\rlap{\textsuperscript{2}}\textit{\textsubscript{n,rms}}\textit{P\textsubscript{tot}}$ to $V\rlap{\textsuperscript{2}}\textit{\textsubscript{n,bjt,tot}}\textit{P\textsubscript{bjt,tot}}$, where $V\rlap{\textsuperscript{2}}\textit{\textsubscript{n,bjt,tot}}$ and $\textit{P\textsubscript{bjt,tot}}$ are the input-referred noise voltage and power consumption of the BJT, respectively. As in the NEF, the input-referred noise voltage of the BJT is calculated using the common-emitter stage. Then, the PEF can be re-expressed as

\begin{align}
	\frac{V\rlap{\textsuperscript{2}}\textit{\textsubscript{n,rms}}\cdot\textit{P\textsubscript{tot}}}{V\rlap{\textsuperscript{2}}\textit{\textsubscript{n,bjt,tot}}\cdot\textit{P\textsubscript{bjt,tot}}} &= \frac{V\rlap{\textsuperscript{2}}\textit{\textsubscript{n,rms}}\cdot\textit{I\textsubscript{DD}}\textit{V\textsubscript{DD}}}{V\rlap{\textsuperscript{2}}\textit{\textsubscript{n,bjt,tot}}\cdot\textit{I\textsubscript{C}}\textit{V\textsubscript{DD,bjt}}}\\
	&= \frac{V\rlap{\textsuperscript{2}}\textit{\textsubscript{n,rms}}\cdot\textit{I\textsubscript{DD}}\textit{V\textsubscript{DD}}}{\frac{\pi\cdot\textit{U\textsubscript{T}}\cdot4kT\cdot{\textit{BW\textsubscript{bjt}}}}{2\textit{I\textsubscript{C}}}\cdot\textit{I\textsubscript{C}}\textit{V\textsubscript{DD,bjt}}}
\end{align}

\noindent
where $\textit{I\textsubscript{C}}$ and $\textit{V\textsubscript{DD,bjt}}$ are the collector bias current and supply voltage of the BJT, respectively. Note that $V\rlap{\textsuperscript{2}}\textit{\textsubscript{n,bjt,tot}}$ is calculated from the common-emitter stage using only the collector shot noise while neglecting the base shot noise, base resistance, and flicker noise. Therefore, $V\rlap{\textsuperscript{2}}\textit{\textsubscript{n,bjt,tot}}$ of Eq. (11) can be expressed as $\frac{\pi\cdot\textit{U\textsubscript{T}}\cdot4kT\cdot{\textit{BW\textsubscript{bjt}}}}{2\textit{I\textsubscript{C}}}$ as presented in \cite{r6}.

\indent
$V\rlap{\textsuperscript{2}}\textit{\textsubscript{n,bjt,tot}}$ is calculated by dividing the collector shot noise by the BJT transconductance and integrating it over the noise bandwidth $\textit{BW\textsubscript{bjt}}\cdot\frac{\pi}{2}$. Also, the front-end bandwidth $\textit{BW\textsubscript{fe}}$ is equalized to the BJT bandwidth $\textit{BW\textsubscript{bjt}}$ for obtaining a normalized efficiency factor over the BJT, meaning $\textit{BW\textsubscript{bjt}} = \textit{BW\textsubscript{fe}}$. Assuming the PEF is expressed using the normalized 1 V supply voltage of the BJT ($\textit{V\textsubscript{DD,bjt}}$ = 1 V) as discussed in \cite{r9}, \cite{r10}, Eq. (12) can be summarized as

\begin{align}
	\frac{V\rlap{\textsuperscript{2}}\textit{\textsubscript{n,rms}}\cdot\textit{I\textsubscript{DD}}\textit{V\textsubscript{DD}}}{\frac{\pi\cdot\textit{U\textsubscript{T}}\cdot4kT\cdot{\textit{BW\textsubscript{fe}}}}{2}\cdot1} &= V\rlap{\textsuperscript{2}}\textit{\textsubscript{n,rms}}\frac{2\textit{I\textsubscript{DD}}\textit{V\textsubscript{DD}}}{\pi\cdot\textit{U\textsubscript{T}}\cdot4kT\cdot{\textit{BW\textsubscript{fe}}}}\\
	&= \text{NEF}\textsuperscript{2}\textit{V\textsubscript{DD}}
\end{align} 

\indent
As shown in the equations above, the PEF is explained as follows. The product of the input-referred noise and power consumption is normalized with respect to the common-emitter stage. Normalization is carried out assuming that the BJT is driven using a 1 V supply voltage ($\textit{V\textsubscript{DD,bjt}}$ = 1 V). The bandwidth is equalized across the front-end and common-emitter stage ($\textit{BW\textsubscript{bjt}} = \textit{BW\textsubscript{fe}}$). Therefore, using the following assumptions of $\textit{V\textsubscript{DD,bjt}}$ = 1 V and $\textit{BW\textsubscript{bjt}} = \textit{BW\textsubscript{fe}}$, Eq. (12) is reduced to Eqs. (13) and (14). In reality, Eqs. (13) and (14) have been widely employed as the FoM for fairly comparing the performance of diverse biopotential recording front-ends, even if the front-ends achieve different $\textit{V\textsubscript{n,rms}}$ over $\textit{BW\textsubscript{fe}}$ while consuming the power of $\textit{I\textsubscript{DD}}\textit{V\textsubscript{DD}}$.

\indent
The procedure of Eqs. (11) and (12) can be applied to either a single-ended common-emitter stage or a differential common-emitter stage by using the BJT noise models in Eq. (9). Assuming that $\textit{V\textsubscript{DD,bjt}} =$ 1 V and $\textit{BW\textsubscript{bjt}} = \textit{BW\textsubscript{fe}}$, the following equations provide the power efficiency factors with respect to a single-ended common-emitter stage and a differential common-emitter stage:

\begin{align}
	\frac{V\rlap{\textsuperscript{2}}\textit{\textsubscript{n,rms}}\cdot\textit{P\textsubscript{tot}}}{V\rlap{\textsuperscript{2}}\textit{\textsubscript{n,single-bjt,tot}}\cdot\textit{P\textsubscript{single-bjt,tot}}} &= \frac{V\rlap{\textsuperscript{2}}\textit{\textsubscript{n,rms}}\cdot\textit{I\textsubscript{DD}}\textit{V\textsubscript{DD}}}{\frac{\pi\cdot\textit{U\textsubscript{T}}\cdot2kT\cdot{\textit{BW\textsubscript{bjt}}}}{2\textit{I\textsubscript{C}}}\cdot\textit{I\textsubscript{C}}\textit{V\textsubscript{DD,bjt}}} \,\,\,\,\,\,\,\, \text{Single-Ended Stage}\\ 
	&= \frac{V\rlap{\textsuperscript{2}}\textit{\textsubscript{n,rms}}\cdot\textit{I\textsubscript{DD}}\textit{V\textsubscript{DD}}}{\frac{\pi\cdot\textit{U\textsubscript{T}}\cdot2kT\cdot\textit{BW\textsubscript{fe}}}{2}\cdot1}\\
	&= V\rlap{\textsuperscript{2}}\textit{\textsubscript{n,rms}}\frac{\textit{I\textsubscript{DD}}\textit{V\textsubscript{DD}}}{\pi\cdot\textit{U\textsubscript{T}}\cdot{kT}\cdot{\textit{BW\textsubscript{fe}}}}\\
	&= NEF\rlap{\textsuperscript{2}}\textit{\textsubscript{single-bjt}}\textit{V\textsubscript{DD}}
\end{align}

\begin{align}
	\frac{V\rlap{\textsuperscript{2}}\textit{\textsubscript{n,rms}}\cdot\textit{P\textsubscript{tot}}}{V\rlap{\textsuperscript{2}}\textit{\textsubscript{n,diff-bjt,tot}}\cdot\textit{P\textsubscript{diff-bjt,tot}}} &= \frac{V\rlap{\textsuperscript{2}}\textit{\textsubscript{n,rms}}\cdot\textit{I\textsubscript{DD}}\textit{V\textsubscript{DD}}}{\frac{\pi\cdot\textit{U\textsubscript{T}}\cdot4kT\cdot{\textit{BW\textsubscript{bjt}}}}{2\textit{I\textsubscript{C}}}\cdot2\textit{I\textsubscript{C}}\textit{V\textsubscript{DD,bjt}}} \,\,\,\,\,\,\,\, \text{Differential Stage}\\
	&= \frac{V\rlap{\textsuperscript{2}}\textit{\textsubscript{n,rms}}\cdot\textit{I\textsubscript{DD}}\textit{V\textsubscript{DD}}}{\frac{\pi\cdot\textit{U\textsubscript{T}}\cdot4kT\cdot\textit{BW\textsubscript{fe}}}{1}\cdot1}\\
	&= V\rlap{\textsuperscript{2}}\textit{\textsubscript{n,rms}}\frac{\textit{I\textsubscript{DD}}\textit{V\textsubscript{DD}}}{\pi\cdot\textit{U\textsubscript{T}}\cdot4kT\cdot{\textit{BW\textsubscript{fe}}}}\\
	&= \frac{1}{2}{NEF\rlap{\textsuperscript{2}}\textit{\textsubscript{diff-bjt}}}\textit{V\textsubscript{DD}}
\end{align}

\noindent
where $\textit{P\textsubscript{single-bjt,tot}}$ and $\textit{P\textsubscript{diff-bjt,tot}}$ mean the power consumption in a single-ended common-emitter stage and a differential common-emitter stage, respectively. Since the bias current of a differential stage consumes two times the current of a single stage, the bias current of Eq. (19) is twice that of Eq. (15). When Eqs. (18) and (22) are defined as $\textit{PEF\textsubscript{single-bjt}}$ and $\textit{PEF\textsubscript{diff-bjt}}$, respectively, $\textit{PEF\textsubscript{single-bjt}}$ and $\textit{PEF\textsubscript{diff-bjt}}$ can be given as follows

\begin{align}
	\textit{PEF\textsubscript{single-bjt}} &= NEF\rlap{\textsuperscript{2}}\textit{\textsubscript{single-bjt}}\textit{V\textsubscript{DD}}\\
	&= 2\text{NEF}\textsuperscript{2}\textit{V\textsubscript{DD}}
\end{align}

\begin{align}
	\textit{PEF\textsubscript{diff-bjt}} &= \frac{1}{2}{NEF\rlap{\textsuperscript{2}}\textit{\textsubscript{diff-bjt}}\textit{V\textsubscript{DD}}}\\
	&= \frac{1}{2}\text{NEF}\textsuperscript{2}\textit{V\textsubscript{DD}}
\end{align}

\noindent
where $\textit{NEF\textsubscript{single-bjt}} = \sqrt{2}\text{NEF}$ and $\textit{NEF\textsubscript{diff-bjt}} = \text{NEF}$. Then, the power efficiency factors across the single and differential stages can be summarized as $\textit{PEF\textsubscript{single-bjt}} = 4\textit{PEF\textsubscript{diff-bjt}}$.\\

\subsection{\textit{PEF Trends}}
\begin{figure}[h]
 	\centering
 	\includegraphics[scale=0.55]{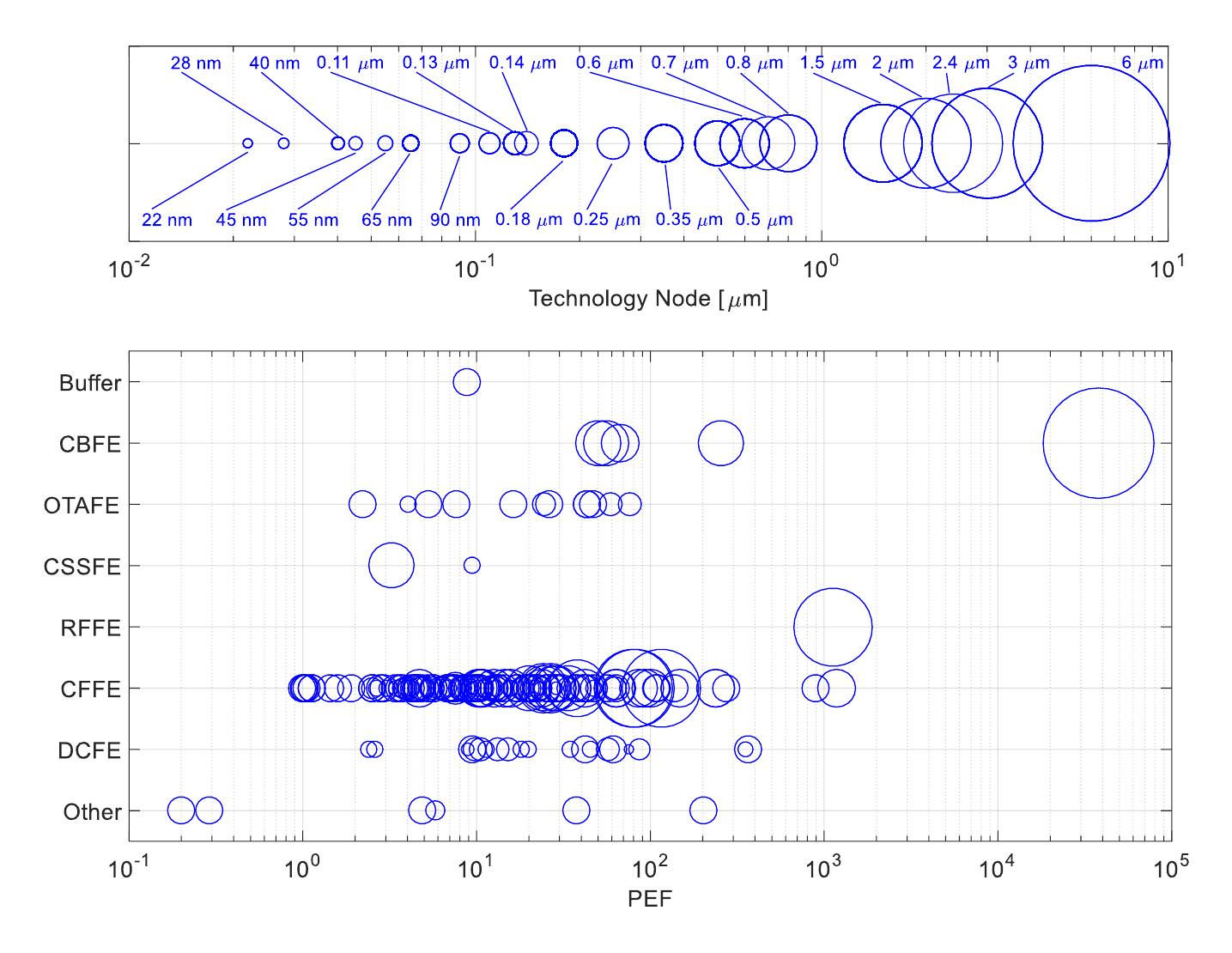}
 	\caption{PEF according to front-end architectures and technology nodes.}
 	\label{Fig. 3}
 \end{figure}

As discussed in equations from Eq. (11) to Eq. (14), the PEF is simply calculated as $\text{NEF}\textsuperscript{2}\textit{V\textsubscript{DD}}$ and has been widely employed as the FoM along with the NEF. Fig. \ref{Fig. 3} presents the overall PEF trends over the last decades. As already discussed in the NEF trends shown in Fig. \ref{Fig. 2}, PEFs noted in Fig. \ref{Fig. 3} also do not reflect all the front-ends shown in Fig. \ref{Fig. 1} due to the lack of information in prior literature. Note that the concept of PEF \cite{r8} was developed in the 2010s, which came out long after the NEF \cite{r6} was introduced. The top of Fig. \ref{Fig. 3} describes the symbol sizes according to the technology nodes ranging from 22 nm to 6 $\mu$m. The bottom of Fig. \ref{Fig. 3} summarizes the overall PEF trends according to the front-end architectures shown in Fig. \ref{Fig. 1}.

\indent
Compared to the NEF trends, the PEFs achieved using the CFFE are mostly distributed from 1 to 100 (Fig. \ref{Fig. 3}), while the NEFs achieved using the CFFE mostly range between 1 to 10 (Fig. \ref{Fig. 2}). In the design of biopotential recording front-ends, lower values of both NEF and PEF indicate better performance in current-noise efficiency and power-noise efficiency, respectively. Therefore, many front-end designers have tried to obtain lower values of both NEF and PEF.

\indent
Considering that the PEF is obtained by squaring the NEF and multiplying it by a supply voltage, a higher PEF is achieved if the NEF is larger than 1, which is not a preferable direction in the front-end design. On the other hand, even if a supply voltage is somewhat higher than 1 V, a lower PEF can be achieved if the NEF is less than 1. Also, for achieving a lower PEF, another way instead of lowering the NEF is to drive the front-end using a sub-1 V supply voltage.

\indent
However, lowering the NEF and supply voltage in a front-end design could be a challenging task considering a reliable circuit operation and performance degradation. For example, the chopper in the CFFE can reduce the input-referred noise and eventually lower the NEF. However, the chopping frequency reduces the input impedance of a front-end, which is related to the CMRR degradation. Also, the DC offsets are modulated by the chopping frequency, thereby requiring additional devices to suppress the modulated offsets and eventually increasing the design area. Likewise, a design technique to improve a specific performance may be associated with other performance degradation.
 
\section{\Large{M}\large{APPING} \large{OF} \Large{T}\large{ECHNOLOGY} \Large{N}\large{ODES}, \Large{N}\large{EF}, \large{AND} \Large{P}\large{EF}}
When using a biopotential recording front-end \textit{in vivo}, weak voltage signals must be processed while minimizing signal contamination, and the heat generated in the circuits must be also minimized to prevent cell damage. Therefore, a front-end must be implemented to have low noise while consuming low power, which means that both NEF and PEF must be also achieved as lower values. In this section, the overall performance summary including the technology nodes, NEF, and PEF is presented for the biopotential recording front-ends developed over the last decades. Then, the performance summary of each front-end architecture is also presented.\\

\subsection{\textit{Mapping of Overall Front-End Architectures}}
When revisiting the equation of $\text{PEF} = \text{NEF}\textsuperscript{2}\textit{V\textsubscript{DD}}$, even if the NEF is reduced thanks to circuit techniques, the PEF could be higher if a higher $\textit{V\textsubscript{DD}}$ is used. It means that the power-noise efficiency does not advance even if the current-noise efficiency improves. Also, even if the NEF is somewhat degraded due to a high bias current or input noise, the PEF could advance if a lower $\textit{V\textsubscript{DD}}$ is used. It means that the power-noise efficiency advances even if the current-noise efficiency is degraded. In fact, the intuitive way to reduce the power consumption of circuits is to use a lower $\textit{V\textsubscript{DD}}$. However, using a lower $\textit{V\textsubscript{DD}}$ could be limited when considering the bias conditions of transistors, dynamic range, etc.

\indent
Ultimately, considering a front-end used \textit{in vivo}, both NEF and PEF must be reduced, which means that both current-noise efficiency and power-noise efficiency must be improved. Therefore, to figure out how well the current-noise efficiency and power-noise efficiency are balanced rather than being biased to one side, $|\text{PEF} - \text{NEF}|$ is employed, which means the difference between NEF and PEF. Note that when the absolute value sign is not used in $|\text{PEF} - \text{NEF}|$, the positive value of $\text{PEF} - \text{NEF}$ means the power-noise efficiency is higher than the current-noise efficiency, and the negative value of $\text{PEF} - \text{NEF}$ means the current-noise efficiency is better than the power-noise efficiency. But eventually, achieving a lower $|\text{PEF} - \text{NEF}|$ is the preferable direction in a front-end design.

\begin{figure}[h]
	\centering
	\includegraphics[scale=0.55]{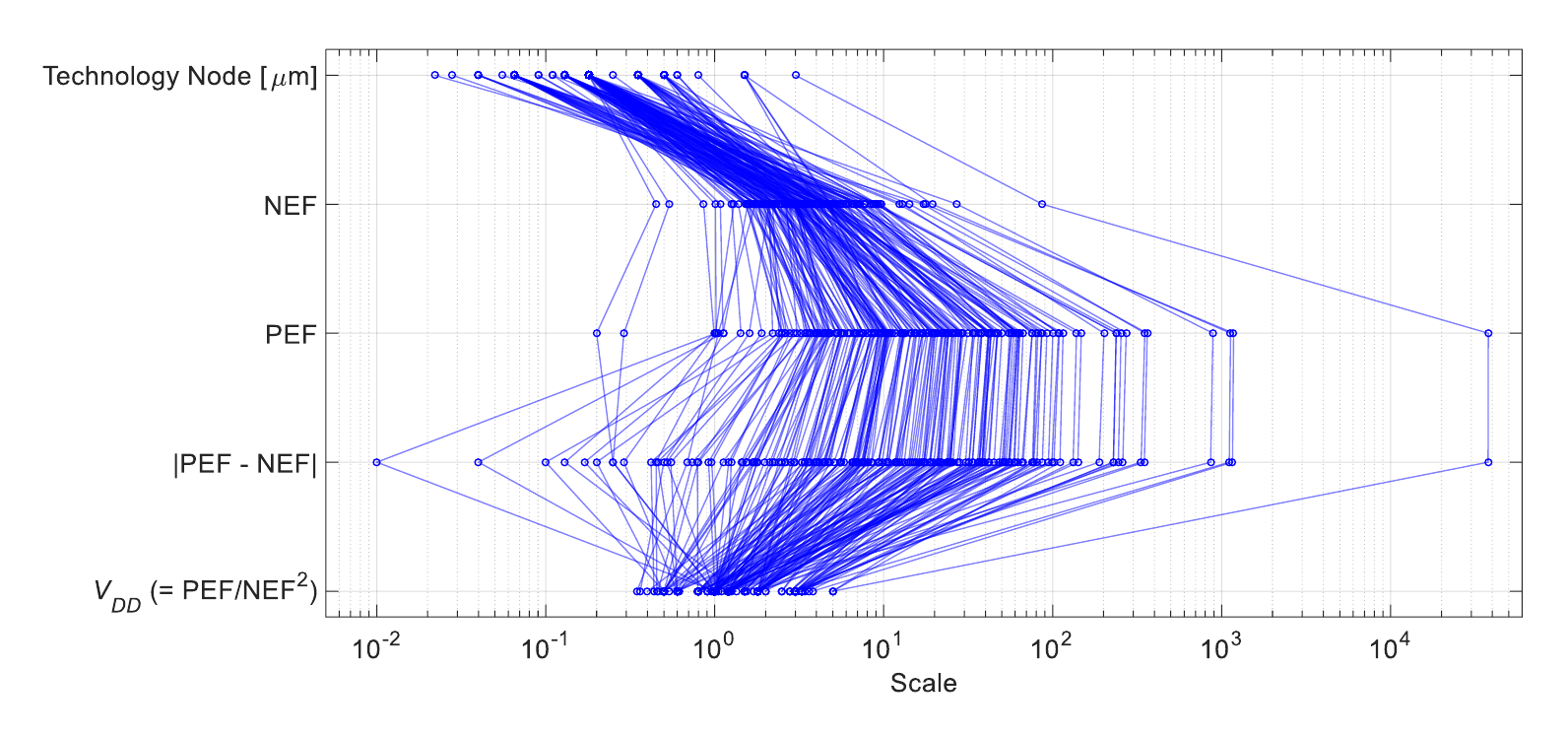}
	\caption{Performance mapping across the overall front-end architectures.}
	\label{Fig. 4}
\end{figure}

\indent
Fig. \ref{Fig. 4} presents the mapping of the front-end performance by using the technology nodes, NEF, and PEF for all architectures shown in Fig. \ref{Fig. 1}. The x-axis indicates a numeric scale. The y-axis includes five categories of the technology node, NEF, PEF, $|\text{PEF} - \text{NEF}|$, and $\textit{V\textsubscript{DD}}$. Note that $\textit{V\textsubscript{DD}}$ is obtained through the reverse calculation using $\text{PEF}/\text{NEF}\textsuperscript{2}$. Each blue line indicates the performance mapping with respect to a single front-end (Fig. \ref{Fig. 4}). In other words, five data points are connected according to the corresponding values held by a front-end. Lowering NEF, PEF, and $|\text{PEF} - \text{NEF}|$ makes the front-end suitable for use in systems where the power and noise must be strictly regulated.

\subsection{\textit{Mapping According to Architectures}}
\begin{figure}[h]
	\centering
	\includegraphics[scale=0.25]{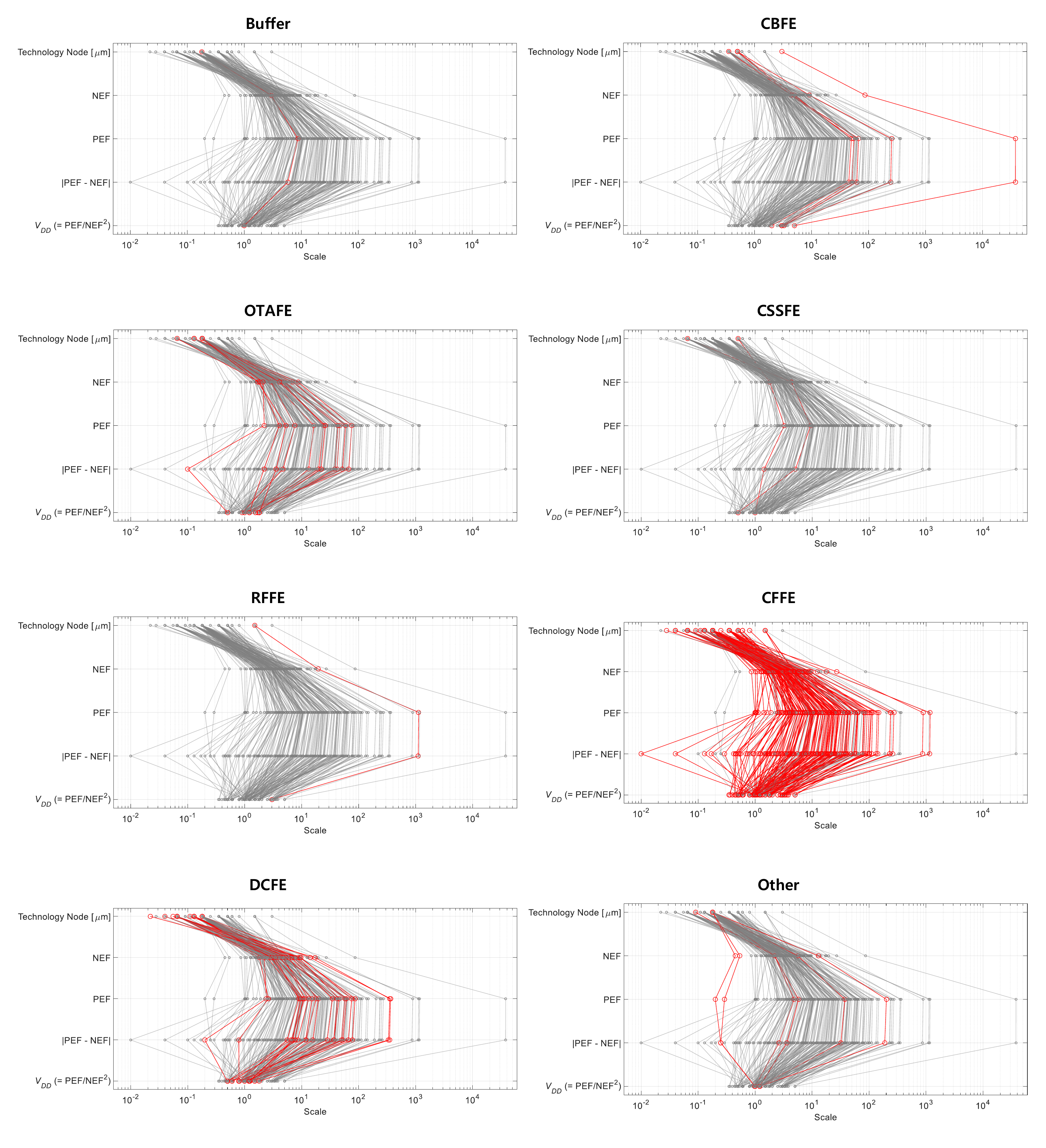}
	\caption{Performance mapping for each front-end architecture.}
	\label{Fig. 5}
\end{figure}

Compared to Fig. \ref{Fig. 4} describing the mapping across the overall architectures shown in Fig. \ref{Fig. 1}, Fig. \ref{Fig. 5} presents the performance mappings depending on the front-end architectures. Note that all blue lines shown in Fig. \ref{Fig. 4} change to gray lines, and red lines are drawn to represent each architecture as shown in Fig. \ref{Fig. 5}. As shown in Figs. \ref{Fig. 1} and \ref{Fig. 5}, among the architectures that constitute biopotential recording front-ends, the CFFE has been overwhelmingly employed, and the DCFE is the second most used architecture. In reality, the number of third and fourth most used architectures shown in Fig. \ref{Fig. 1} does not reflect completely the number of architectures shown in Fig. \ref{Fig. 5} due to the following reason. Many research papers presenting the front-ends do not specify either NEF or supply voltage, or neither. If the NEF and supply voltage are noted but the PEF is not noted in prior research papers, the PEF is calculated using $\text{NEF}\textsuperscript{2}\textit{V\textsubscript{DD}}$ in this article.

\indent
In particular, the diverse circuit techniques for reducing the input-referred noise while reducing the supply voltage have been developed based on the CFFE, resulting in lower values in NEF, PEF, and $|\text{PEF} - \text{NEF}|$. Also, while many CFFEs have been developed using a wide range of technology nodes, most DCFEs are developed using scaled-down technology nodes than those of CFFEs (Fig. \ref{Fig. 5}).

\subsection{\textit{Noise Analysis of MMRS}}
\begin{figure}[h]
	\centering
	\includegraphics[scale=0.6]{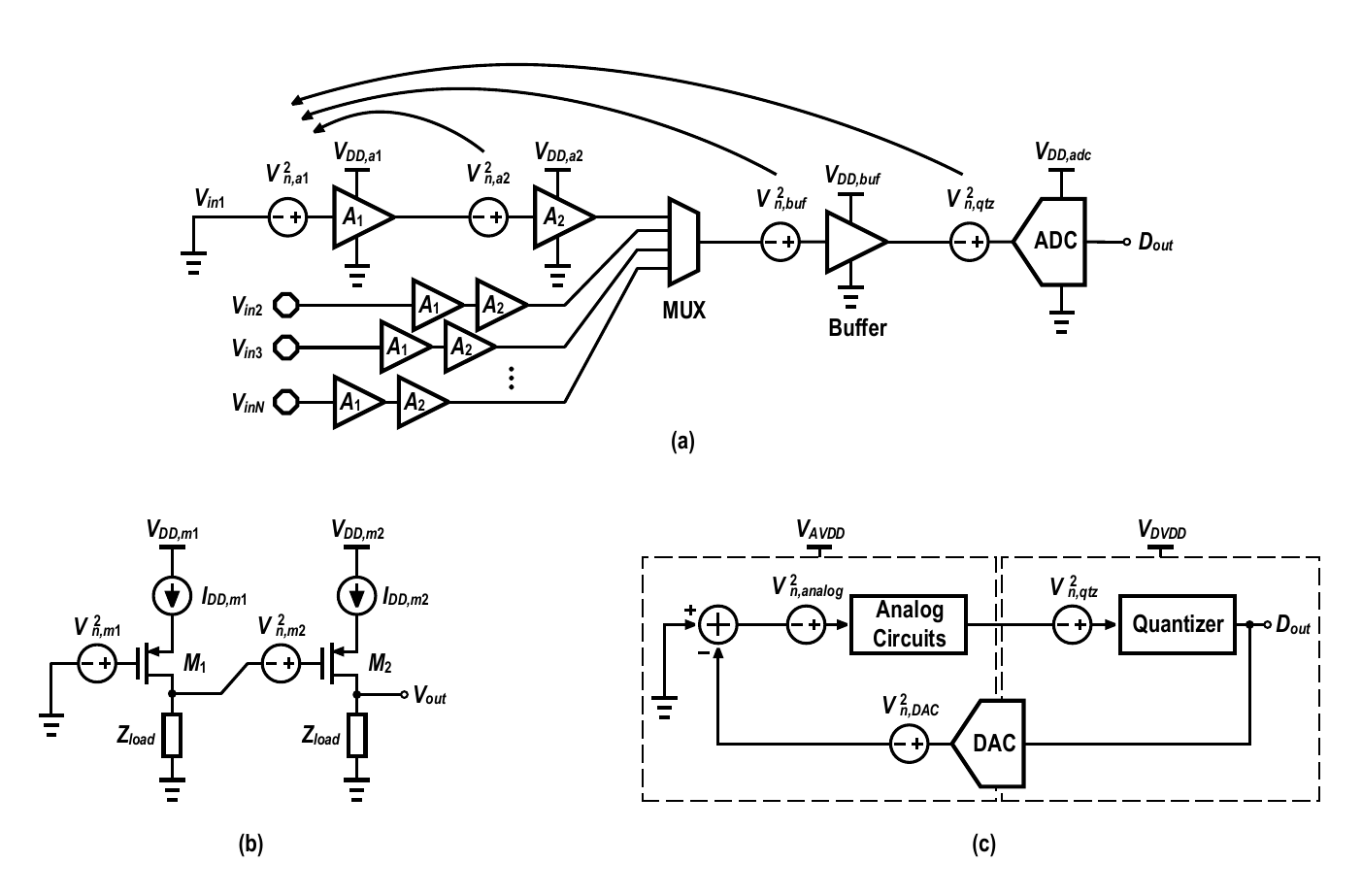}
	\caption{Simplified circuit diagrams including the noise sources: (a) multi-channel multi-stage recording system, (b) two-stage amplifier with different supply voltages, and (c) direct conversion front-end that performs delta-sigma operations.}
	\label{Fig. 6}
\end{figure}

The multi-channel multi-stage recording system (MMRS) sharing a single ADC is a widely used system architecture for an area-efficient design. Assuming that the input signal is digitized after two amplification stages, Fig. \ref{Fig. 6}(a) shows the general architecture of the MMRS with noise sources. The input signal is processed through a recording chain from the first amplification stage $A\textsubscript{1}$ to the ADC. Even if the overall input-referred noise voltage is dominated by $A\textsubscript{1}$ in a whole recording chain, depending on the gain of the preceding stages, the input noise could be slightly influenced by the subsequent stages, e.g., the second amplification stage $A\textsubscript{2}$, buffer, and ADC. If $A\textsubscript{2}$ has sufficient capability for driving the ADC, the buffer can be left out. Considering all circuit blocks shown in Fig. \ref{Fig. 6}(a), the overall input-referred noise voltage $V\rlap{\textsuperscript{2}}\textit{\textsubscript{n,rms}}$ with respect to a single recording chain is obtained as

\begin{align}
	V\rlap{\textsuperscript{2}}\textit{\textsubscript{n,rms}} &= V\rlap{\textsuperscript{2}}\textit{\textsubscript{n,a}}\textsubscript{1} + \frac{V\rlap{\textsuperscript{2}}\textit{\textsubscript{n,a}}\textsubscript{2}}{G\rlap{\textsuperscript{2}}\textsubscript{1}} + \frac{V\rlap{\textsuperscript{2}}\textit{\textsubscript{n,buf}}}{\big(G\textsubscript{1}G\textsubscript{2}\big)^2} + \frac{V\rlap{\textsuperscript{2}}\textit{\textsubscript{n,qtz}}}{\big(G\textsubscript{1}G\textsubscript{2}\big)^2}
\end{align}

\noindent
where $G\textsubscript{1}$ and $G\textsubscript{2}$ are voltage gains of $A\textsubscript{1}$ and $A\textsubscript{2}$, respectively. The noise sources shown in Fig. \ref{Fig. 6}(a) indicate the input-referred noise voltage of each circuit block as follows: (1) the first amplification stage $A\textsubscript{1}$, $V\rlap{\textsuperscript{2}}\textit{\textsubscript{n,a}}\textsubscript{1}$; (2) the second amplification stage $A\textsubscript{2}$, $V\rlap{\textsuperscript{2}}\textit{\textsubscript{n,a}}\textsubscript{2}$; (3) the ADC driving buffer, $V\rlap{\textsuperscript{2}}\textit{\textsubscript{n,buf}}$; and (4) the ADC, $V\rlap{\textsuperscript{2}}\textit{\textsubscript{n,qtz}}$. Each noise source is divided by the gain of the preceding stages to obtain the overall input-referred noise voltage in a recording chain. Therefore, assuming that the buffer gain is set to unity, $V\rlap{\textsuperscript{2}}\textit{\textsubscript{n,qtz}}$ is divided by $\big(G\textsubscript{1}G\textsubscript{2}\big)^2$. Eq. (27) can be approximated as $V\rlap{\textsuperscript{2}}\textit{\textsubscript{n,a}}\textsubscript{1}$ if $G\textsubscript{1}G\textsubscript{2}$ and $G\textsubscript{1}$ are sufficiently high, e.g., $G\textsubscript{1} = 100$ and $G\textsubscript{2} = 10$. Assuming that $A\textsubscript{1}$ and $A\textsubscript{2}$ are developed as the CFFE, the way to lower the input noise is to increase each gain, and in particular, increasing $G\textsubscript{1}$ can effectively reduce $V\rlap{\textsuperscript{2}}\textit{\textsubscript{n,rms}}$. However, increasing $G\textsubscript{1}$ leads to a large design area of $A\textsubscript{1}$, which encounters a trade-off between the noise and area. Note that each circuit block shown in Fig. \ref{Fig. 6}(a) draws the current as follows: $\textit{I\textsubscript{DD,a}}\textsubscript{1}$ from $A\textsubscript{1}$, $\textit{I\textsubscript{DD,a}}\textsubscript{2}$ from $A\textsubscript{2}$, $\textit{I\textsubscript{DD,buf}}$ from the buffer, and $\textit{I\textsubscript{DD,adc}}$ from the ADC.

\indent
Assuming a single recording chain in Fig. \ref{Fig. 6}(a) achieves the bandwidth $\textit{BW\textsubscript{ch}}$ while consuming the total bias current $\textit{I\textsubscript{DD,ch}}$ ($= \textit{I\textsubscript{DD,a}}\textsubscript{1} + \textit{I\textsubscript{DD,a}}\textsubscript{2} + \textit{I\textsubscript{DD,buf}} + \textit{I\textsubscript{DD,adc}}$), the input-referred noise voltage of a single chain can be compared to that of a single-ended common-emitter stage as follows

\begin{align}
	\frac{V\textit{\textsubscript{n,rms}}}{V\textit{\textsubscript{n,single-bjt,tot}}} &= \sqrt{V\rlap{\textsuperscript{2}}\textit{\textsubscript{n,a}}\textsubscript{1} + \frac{V\rlap{\textsuperscript{2}}\textit{\textsubscript{n,a}}\textsubscript{2}}{G\rlap{\textsuperscript{2}}\textsubscript{1}} + \frac{V\rlap{\textsuperscript{2}}\textit{\textsubscript{n,buf}}}{\big(G\textsubscript{1}G\textsubscript{2}\big)^2} + \frac{V\rlap{\textsuperscript{2}}\textit{\textsubscript{n,qtz}}}{\big(G\textsubscript{1}G\textsubscript{2}\big)^2}}\Bigg/{\sqrt{\frac{2kT\cdot\textit{U\textsubscript{T}}}{\textit{I\textsubscript{C}}}\cdot{\textit{BW\textsubscript{bjt}}}\cdot\frac{\pi}{2}}}
\end{align}

\noindent
where $V\textit{\textsubscript{n,single-bjt,tot}}$ and $V\textit{\textsubscript{n,rms}}$ are derived from Eqs. (9) and (27), respectively. When equalizing the bandwidth and total bias current of the BJT to those of a single recording chain, $\textit{I\textsubscript{C}} = \textit{I\textsubscript{DD,ch}}$ and $\textit{BW\textsubscript{bjt}} = \textit{BW\textsubscript{ch}}$. Therefore, Eq. (28) can be rewritten as

\begin{align}
	\frac{V\textit{\textsubscript{n,rms}}}{V\textit{\textsubscript{n,single-bjt,tot}}} &= \sqrt{\Bigg[V\rlap{\textsuperscript{2}}\textit{\textsubscript{n,a}}\textsubscript{1} + \frac{V\rlap{\textsuperscript{2}}\textit{\textsubscript{n,a}}\textsubscript{2}}{G\rlap{\textsuperscript{2}}\textsubscript{1}} + \frac{V\rlap{\textsuperscript{2}}\textit{\textsubscript{n,buf}}}{\big(G\textsubscript{1}G\textsubscript{2}\big)^2} + \frac{V\rlap{\textsuperscript{2}}\textit{\textsubscript{n,qtz}}}{\big(G\textsubscript{1}G\textsubscript{2}\big)^2}\Bigg]\frac{\textit{I\textsubscript{DD,ch}}}{\pi\cdot\textit{U\textsubscript{T}}\cdot{kT}\cdot\textit{BW\textsubscript{ch}}}}
\end{align}

\indent
As calculated in Eqs. (28) and (29), the noise voltage of a single chain can be also compared to that of a differential common-emitter stage by using Eq. (9). After equalizing the bandwidth and total bias current across the single chain and differential BJT, the following equation is obtained:

\begin{align}
	\frac{V\textit{\textsubscript{n,rms}}}{V\textit{\textsubscript{n,diff-bjt,tot}}} &= \sqrt{\Bigg[V\rlap{\textsuperscript{2}}\textit{\textsubscript{n,a}}\textsubscript{1} + \frac{V\rlap{\textsuperscript{2}}\textit{\textsubscript{n,a}}\textsubscript{2}}{G\rlap{\textsuperscript{2}}\textsubscript{1}} + \frac{V\rlap{\textsuperscript{2}}\textit{\textsubscript{n,buf}}}{\big(G\textsubscript{1}G\textsubscript{2}\big)^2} + \frac{V\rlap{\textsuperscript{2}}\textit{\textsubscript{n,qtz}}}{\big(G\textsubscript{1}G\textsubscript{2}\big)^2}\Bigg]\frac{\textit{I\textsubscript{DD,ch}}}{\pi\cdot\textit{U\textsubscript{T}}\cdot{2kT}\cdot\textit{BW\textsubscript{ch}}}}
\end{align}

\indent
The product of the noise voltage and power consumption of a single chain can be compared to that of the BJT. All supply voltages used in a single chain are equal to $\textit{V\textsubscript{DD,ch}}$. Assuming that the BJT is driven using a 1-V supply voltage ($\textit{V\textsubscript{DD,bjt}} =$ 1 V) and $\textit{BW\textsubscript{bjt}} = \textit{BW\textsubscript{ch}}$, the power-noise efficiency between a single recording chain and a single-ended common-emitter stage is given by

\begin{align}
	\frac{V\rlap{\textsuperscript{2}}\textit{\textsubscript{n,rms}}\cdot\textit{I\textsubscript{DD,ch}}\textit{V\textsubscript{DD,ch}}}{V\rlap{\textsuperscript{2}}\textit{\textsubscript{n,single-bjt,tot}}\cdot\textit{I\textsubscript{C}}\textit{V\textsubscript{DD,bjt}}} &= \frac{\Bigg[V\rlap{\textsuperscript{2}}\textit{\textsubscript{n,a}}\textsubscript{1} + \frac{V\rlap{\textsuperscript{2}}\textit{\textsubscript{n,a}}\textsubscript{2}}{G\rlap{\textsuperscript{2}}\textsubscript{1}} + \frac{V\rlap{\textsuperscript{2}}\textit{\textsubscript{n,buf}}}{\big(G\textsubscript{1}G\textsubscript{2}\big)^2} + \frac{V\rlap{\textsuperscript{2}}\textit{\textsubscript{n,qtz}}}{\big(G\textsubscript{1}G\textsubscript{2}\big)^2}\Bigg]\textit{I\textsubscript{DD,ch}}\textit{V\textsubscript{DD,ch}}}{\frac{\pi\cdot\textit{U\textsubscript{T}}\cdot2kT\cdot{\textit{BW\textsubscript{bjt}}}}{2\textit{I\textsubscript{C}}}\textit{I\textsubscript{C}}\textit{V\textsubscript{DD,bjt}}}\\
	&= \frac{\Bigg[V\rlap{\textsuperscript{2}}\textit{\textsubscript{n,a}}\textsubscript{1} + \frac{V\rlap{\textsuperscript{2}}\textit{\textsubscript{n,a}}\textsubscript{2}}{G\rlap{\textsuperscript{2}}\textsubscript{1}} + \frac{V\rlap{\textsuperscript{2}}\textit{\textsubscript{n,buf}}}{\big(G\textsubscript{1}G\textsubscript{2}\big)^2} + \frac{V\rlap{\textsuperscript{2}}\textit{\textsubscript{n,qtz}}}{\big(G\textsubscript{1}G\textsubscript{2}\big)^2}\Bigg]\textit{I\textsubscript{DD,ch}}\textit{V\textsubscript{DD,ch}}}{\pi\cdot\textit{U\textsubscript{T}}\cdot{kT}\cdot{\textit{BW\textsubscript{ch}}}}
\end{align}

\indent
If the circuit blocks in a single chain are driven using different supply voltages to optimize the power consumption, the total power consumption of a single chain must be expressed using the individual supply voltage of each circuit block. Therefore, Eq. (32) can be rewritten as

\begin{align}
	\frac{V\rlap{\textsuperscript{2}}\textit{\textsubscript{n,rms}}\cdot\text{Channel Power}}{V\rlap{\textsuperscript{2}}\textit{\textsubscript{n,single-bjt,tot}}\cdot\textit{I\textsubscript{C}}\textit{V\textsubscript{DD,bjt}}} &= \frac{\Bigg[V\rlap{\textsuperscript{2}}\textit{\textsubscript{n,a}}\textsubscript{1} + \frac{V\rlap{\textsuperscript{2}}\textit{\textsubscript{n,a}}\textsubscript{2}}{G\rlap{\textsuperscript{2}}\textsubscript{1}} + \frac{V\rlap{\textsuperscript{2}}\textit{\textsubscript{n,buf}}}{\big(G\textsubscript{1}G\textsubscript{2}\big)^2} + \frac{V\rlap{\textsuperscript{2}}\textit{\textsubscript{n,qtz}}}{\big(G\textsubscript{1}G\textsubscript{2}\big)^2}\Bigg]\big(\textit{P\textsubscript{a}}\textsubscript{1} + \textit{P\textsubscript{a}}\textsubscript{2} + \textit{P\textsubscript{buf}} + \textit{P\textsubscript{adc}}\big)}{\pi\cdot\textit{U\textsubscript{T}}\cdot{kT}\cdot{\textit{BW\textsubscript{ch}}}}
\end{align}

\noindent
where $\textit{P\textsubscript{a}}\textsubscript{1} = \textit{I\textsubscript{DD,a}}\textsubscript{1}\textit{V\textsubscript{DD,a}}\textsubscript{1}$, $\textit{P\textsubscript{a}}\textsubscript{2} = \textit{I\textsubscript{DD,a}}\textsubscript{2}\textit{V\textsubscript{DD,a}}\textsubscript{2}$, $\textit{P\textsubscript{buf}} = \textit{I\textsubscript{DD,buf}}\textit{V\textsubscript{DD,buf}}$, and $\textit{P\textsubscript{adc}} = \textit{I\textsubscript{DD,adc}}\textit{V\textsubscript{DD,adc}}$. Also, by expanding Eq. (33) in a differential way while using Eq. (9) and $\textit{V\textsubscript{DD,bjt}} =$ 1 V, the power-noise efficiency between a single chain and a differential common-emitter stage can be obtained as

\begin{align}
	\frac{V\rlap{\textsuperscript{2}}\textit{\textsubscript{n,rms}}\cdot\text{Channel Power}}{V\rlap{\textsuperscript{2}}\textit{\textsubscript{n,diff-bjt,tot}}\cdot2\textit{I\textsubscript{C}}\textit{V\textsubscript{DD,bjt}}} &=
	\frac{\Bigg[V\rlap{\textsuperscript{2}}\textit{\textsubscript{n,a}}\textsubscript{1} + \frac{V\rlap{\textsuperscript{2}}\textit{\textsubscript{n,a}}\textsubscript{2}}{G\rlap{\textsuperscript{2}}\textsubscript{1}} + \frac{V\rlap{\textsuperscript{2}}\textit{\textsubscript{n,buf}}}{\big(G\textsubscript{1}G\textsubscript{2}\big)^2} + \frac{V\rlap{\textsuperscript{2}}\textit{\textsubscript{n,qtz}}}{\big(G\textsubscript{1}G\textsubscript{2}\big)^2}\Bigg]\big(\textit{P\textsubscript{a}}\textsubscript{1} + \textit{P\textsubscript{a}}\textsubscript{2} + \textit{P\textsubscript{buf}} + \textit{P\textsubscript{adc}}\big)}{\pi\cdot\textit{U\textsubscript{T}}\cdot{4kT}\cdot{\textit{BW\textsubscript{ch}}}}
\end{align}

\singlespacing
In the calculation of Eq. (34), the bias current of the differential BJT is twice that of the single BJT shown in Eq. (31), therefore, the bias current of $2\textit{I\textsubscript{C}}$ is used in the denominator of Eq. (34). As shown in Eqs. (33) and (34), low noise and low power consumption are required to achieve better power-noise efficiency. To accomplish this goal, the MMRS shown in Fig. \ref{Fig. 6}(a) can be designed as follows: (1) $G\textsubscript{1}$ and $G\textsubscript{2}$ need to be high enough to ignore the noise contributed by the subsequent stages; (2) for optimizing the trade-off between the noise and power in $A\textsubscript{1}$, $\textit{I\textsubscript{DD,a}}\textsubscript{1}$ needs to be greatly increased while maintaining a low $\textit{V\textsubscript{DD,a}}\textsubscript{1}$; and (3) the subsequent stages after $A\textsubscript{1}$ are driven using lower bias currents and higher supply voltages.

\indent
Most biopotential recording front-ends can be designed employing a two-stage amplifier shown in Fig. \ref{Fig. 6}(b), e.g., $A\textsubscript{1}$ and $A\textsubscript{2}$ can be designed as a two-stage amplifier. For simplicity, the two-stage amplifier shown in Fig. \ref{Fig. 6}(b) is drawn as a single-ended structure, even if the practical amplifier is designed differentially. If the circuit blocks constituting the front-ends dominate the overall noise voltage, those blocks can be designed while using individually different supply voltages to optimize the noise and power consumption \cite{r10}, (e.g., the low $\textit{V\textsubscript{DD,m}}\textsubscript{1}$ and high $\textit{I\textsubscript{DD,m}}\textsubscript{1}$ are employed for the input stage $M\textsubscript{1}$, and the high $\textit{V\textsubscript{DD,m}}\textsubscript{2}$ and low $\textit{I\textsubscript{DD,m}}\textsubscript{2}$ are employed for the output stage $M\textsubscript{2}$ as shown in Fig. \ref{Fig. 6}(b)). Generally, $M\textsubscript{1}$ carries a large $\textit{I\textsubscript{DD,m}}\textsubscript{1}$ for reducing the thermal noise while having a large area for the low flicker noise. On the other hand, $M\textsubscript{2}$ can be designed to carry a small $\textit{I\textsubscript{DD,m}}\textsubscript{2}$ since the noise of $M\textsubscript{2}$ can be negligible when dividing it by a high gain of $M\textsubscript{1}$.

\indent
The equations above are obtained assuming that the buffer and ADC are assigned to only a single channel, e.g, the buffer and ADC digitize only $\textit{V\textsubscript{in}}\textsubscript{1}$ by sustaining the multiplexer (MUX) fixed to $\textit{V\textsubscript{in}}\textsubscript{1}$ in Fig. \ref{Fig. 6}(a). This means that $\textit{I\textsubscript{DD,buf}}$ and $\textit{I\textsubscript{DD,adc}}$ are set to the extent that only a single channel is driven. On the other hand, if an ADC is designed to perform Nyquist-rate operations, such as a successive approximation register ADC, a wider bandwidth of the ADC and buffer is required to access all multiple channels. Accordingly, $\textit{I\textsubscript{DD,buf}}$ and $\textit{I\textsubscript{DD,adc}}$ must increase to widen the bandwidth as the number of channels increases. Therefore, in the buffer and ADC in Fig. \ref{Fig. 6}(a), the effective current consumption per single channel needs to be expressed considering the number of total channels:

\begin{align}
	\textit{I\textsubscript{DD,buf,eff}} = \frac{\textit{I\textsubscript{DD,buf}}}{N}
	\,\,\,\,\,\,\,
	\text{and}
	\,\,\,\,\,\,\,
	\textit{I\textsubscript{DD,adc,eff}} = \frac{\textit{I\textsubscript{DD,adc}}}{N}
\end{align}

\noindent
where $\textit{I\textsubscript{DD,buf}}$ and $\textit{I\textsubscript{DD,adc}}$ are the total bias current consumed by the buffer and ADC, respectively, to access all channels. $\textit{I\textsubscript{DD,buf,eff}}$ and $\textit{I\textsubscript{DD,adc,eff}}$ refer to the effective current consumption per single channel of the buffer and ADC, respectively, considering the total number of channels $N$. Applying the effective current consumption per single channel to Eqs. (29) and (33), current-noise efficiency and power-noise efficiency with respect to a single-ended common-emitter stage are modified as

\begin{align}
	\frac{V\textit{\textsubscript{n,rms}}}{V\textit{\textsubscript{n,single-bjt,tot}}} &= \sqrt{V\rlap{\textsuperscript{2}}\textit{\textsubscript{n,a}}\textsubscript{1} + \frac{V\rlap{\textsuperscript{2}}\textit{\textsubscript{n,a}}\textsubscript{2}}{G\rlap{\textsuperscript{2}}\textsubscript{1}} + \frac{V\rlap{\textsuperscript{2}}\textit{\textsubscript{n,buf}}}{\big(G\textsubscript{1}G\textsubscript{2}\big)^2} + \frac{V\rlap{\textsuperscript{2}}\textit{\textsubscript{n,qtz}}}{\big(G\textsubscript{1}G\textsubscript{2}\big)^2}}\nonumber\\
	&\,\,\,\,\,\,\,\,\,\times
	\sqrt{\frac{\textit{I\textsubscript{DD,a}}\textsubscript{1} + \textit{I\textsubscript{DD,a}}\textsubscript{2} + \frac{\textit{I\textsubscript{DD,buf}}}{N} + \frac{\textit{I\textsubscript{DD,adc}}}{N}}{\pi\cdot\textit{U\textsubscript{T}}\cdot{kT}\cdot\textit{BW\textsubscript{ch}}}}	
\end{align}

\begin{align}
	\frac{V\rlap{\textsuperscript{2}}\textit{\textsubscript{n,rms}}\cdot\textit{P\textsubscript{ch,eff}}}{V\rlap{\textsuperscript{2}}\textit{\textsubscript{n,single-bjt,tot}}\cdot\textit{I\textsubscript{C}}\textit{V\textsubscript{DD,bjt}}} &= \Bigg[V\rlap{\textsuperscript{2}}\textit{\textsubscript{n,a}}\textsubscript{1} + \frac{V\rlap{\textsuperscript{2}}\textit{\textsubscript{n,a}}\textsubscript{2}}{G\rlap{\textsuperscript{2}}\textsubscript{1}} + \frac{V\rlap{\textsuperscript{2}}\textit{\textsubscript{n,buf}}}{\big(G\textsubscript{1}G\textsubscript{2}\big)^2} + \frac{V\rlap{\textsuperscript{2}}\textit{\textsubscript{n,qtz}}}{\big(G\textsubscript{1}G\textsubscript{2}\big)^2}\Bigg]\nonumber\\
	&\,\,\,\,\,\,\,\,\,\times
	\frac{\textit{P\textsubscript{ch,eff}}}{\pi\cdot\textit{U\textsubscript{T}}\cdot{kT}\cdot{\textit{BW\textsubscript{ch}}}}
\end{align}

\noindent
where $\textit{V\textsubscript{DD,bjt}} =$ 1 V and $\textit{BW\textsubscript{ch}}$ is the overall bandwidth from $A\textsubscript{1}$ to the ADC while consuming the effective channel power $\textit{P\textsubscript{ch,eff}}$. Also, using Eqs. (30) and (34), the effective current consumption in Eq. (35) can be extended to a differential common-emitter stage. Considering the total number of channels $N$ and the supply voltage of each block, $\textit{P\textsubscript{ch,eff}}$ is expressed as

\begin{align}
	\textit{P\textsubscript{ch,eff}} = \textit{I\textsubscript{DD,a}}\textsubscript{1}\textit{V\textsubscript{DD,a}}\textsubscript{1} + \textit{I\textsubscript{DD,a}}\textsubscript{2}\textit{V\textsubscript{DD,a}}\textsubscript{2} + \frac{\textit{I\textsubscript{DD,buf}}}{N}\textit{V\textsubscript{DD,buf}} + \frac{\textit{I\textsubscript{DD,adc}}}{N}\textit{V\textsubscript{DD,adc}}
\end{align}

\subsection{\textit{Noise Analysis of DCFE}}
Similar to the method for optimizing circuit design in terms of noise and power consumption as discussed in Figs. \ref{Fig. 6}(a) and (b), a direct conversion front-end can be also designed using different supply voltages for analog and digital domains as shown in Fig. \ref{Fig. 6}(c). Generally, the digital circuits can be driven using a low supply voltage $\textit{V\textsubscript{DVDD}}$ while the analog circuits are driven using a relatively high supply voltage $\textit{V\textsubscript{AVDD}}$ to ensure the operating point and dynamic range \cite{r11}, \cite{r12}. Assuming that the gain of the analog circuits in Fig. \ref{Fig. 6}(c) is $\textit{G\textsubscript{analog}}$, the overall noise voltage $V\rlap{\textsuperscript{2}}\textit{\textsubscript{n,x}}$ at the quantizer input node is expressed as

\begin{align}
	V\rlap{\textsuperscript{2}}\textit{\textsubscript{n,x}} &=
	G\rlap{\textsuperscript{2}}\textit{\textsubscript{analog}}\big(V\rlap{\textsuperscript{2}}\textit{\textsubscript{n,analog}} + V\rlap{\textsuperscript{2}}\textit{\textsubscript{n,DAC}}\big) + V\rlap{\textsuperscript{2}}\textit{\textsubscript{n,qtz}}
\end{align}

\noindent
where $V\rlap{\textsuperscript{2}}\textit{\textsubscript{n,analog}}$ is the input noise voltage of the analog circuits, $V\rlap{\textsuperscript{2}}\textit{\textsubscript{n,DAC}}$ is the output noise voltage of the digital-to-analog converter (DAC), and $V\rlap{\textsuperscript{2}}\textit{\textsubscript{n,qtz}}$ is the quantization noise voltage in Fig. \ref{Fig. 6}(c). After solving the noise transfer function of Fig. \ref{Fig. 6}(c) and applying it to $V\rlap{\textsuperscript{2}}\textit{\textsubscript{n,x}}$ appearing at the quantizer input node, the output noise voltage $V\rlap{\textsuperscript{2}}\textit{\textsubscript{n,Dout}}$ is expressed as

\begin{align}
	V\rlap{\textsuperscript{2}}\textit{\textsubscript{n,Dout}} &=
	\Big[G\rlap{\textsuperscript{2}}\textit{\textsubscript{analog}}\big(V\rlap{\textsuperscript{2}}\textit{\textsubscript{n,analog}} + V\rlap{\textsuperscript{2}}\textit{\textsubscript{n,DAC}}\big) + V\rlap{\textsuperscript{2}}\textit{\textsubscript{n,qtz}}\Big]\bigg(\frac{1}{1+\textit{G\textsubscript{analog}}}\bigg)^2
\end{align}

\indent
To conduct a fair noise comparison for different front-ends, an output noise voltage needs to be divided by the conversion gain as conducted in Eq. (27). Therefore, $V\rlap{\textsuperscript{2}}\textit{\textsubscript{n,Dout}}$ is divided by the signal transfer function of Fig. \ref{Fig. 6}(c). Then, the input-referred noise voltage $V\rlap{\textsuperscript{2}}\textit{\textsubscript{n,DCFE,in}}$ of a direct conversion front-end is obtained as follows

\begin{align}
	V\rlap{\textsuperscript{2}}\textit{\textsubscript{n,DCFE,in}} &= \Bigg[\Big[G\rlap{\textsuperscript{2}}\textit{\textsubscript{analog}}\big(V\rlap{\textsuperscript{2}}\textit{\textsubscript{n,analog}} + V\rlap{\textsuperscript{2}}\textit{\textsubscript{n,DAC}}\big) + V\rlap{\textsuperscript{2}}\textit{\textsubscript{n,qtz}}\Big]\bigg(\frac{1}{1+\textit{G\textsubscript{analog}}}\bigg)^2\Bigg]\Bigg/\bigg(\frac{\textit{G\textsubscript{analog}}}{1+\textit{G\textsubscript{analog}}}\bigg)^2\\
	&= \Big[G\rlap{\textsuperscript{2}}\textit{\textsubscript{analog}}\big(V\rlap{\textsuperscript{2}}\textit{\textsubscript{n,analog}} + V\rlap{\textsuperscript{2}}\textit{\textsubscript{n,DAC}}\big) + V\rlap{\textsuperscript{2}}\textit{\textsubscript{n,qtz}}\Big]\bigg({\frac{1}{\textit{G\textsubscript{analog}}}}\bigg)^2\\
	&= V\rlap{\textsuperscript{2}}\textit{\textsubscript{n,analog}} + V\rlap{\textsuperscript{2}}\textit{\textsubscript{n,DAC}} + V\rlap{\textsuperscript{2}}\textit{\textsubscript{n,qtz}}\frac{1}{G\rlap{\textsuperscript{2}}\textit{\textsubscript{analog}}}
\end{align}

\indent
As seen in Eq. (43), the overall input-referred noise voltage of a DCFE can be obtained by dividing the quantization noise by the gain of the analog circuits and adding it to the input noise sources of the analog domain shown in Fig. \ref{Fig. 6}(c). Assuming that $G\rlap{\textsuperscript{2}}\textit{\textsubscript{analog}}$ is high enough, the input noise voltage $V\rlap{\textsuperscript{2}}\textit{\textsubscript{n,DCFE,in}}$ can be approximated as $V\rlap{\textsuperscript{2}}\textit{\textsubscript{n,analog}} + V\rlap{\textsuperscript{2}}\textit{\textsubscript{n,DAC}}$. Also, as calculated in Eqs. (15) and (19), the power-noise efficiency of a DCFE can be compared to different front-ends using the product of the noise voltage and power consumption as follows

\begin{align}
	V\rlap{\textsuperscript{2}}\textit{\textsubscript{n,DCFE,in}}\cdot\textit{P\textsubscript{DCFE}} &= \bigg(V\rlap{\textsuperscript{2}}\textit{\textsubscript{n,analog}} + V\rlap{\textsuperscript{2}}\textit{\textsubscript{n,DAC}} + V\rlap{\textsuperscript{2}}\textit{\textsubscript{n,qtz}}\frac{1}{G\rlap{\textsuperscript{2}}\textit{\textsubscript{analog}}}\bigg)\big(\textit{I\textsubscript{AVDD}}\textit{V\textsubscript{AVDD}} + \textit{I\textsubscript{DVDD}}\textit{V\textsubscript{DVDD}}\big)
\end{align}

\noindent
where $\textit{P\textsubscript{DCFE}}$ is the total power consumption of a DCFE shown in Fig. \ref{Fig. 6}(c). $\textit{I\textsubscript{AVDD}}$ and $\textit{I\textsubscript{DVDD}}$ are the overall bias currents consumed by the analog domain and digital domain, respectively. Generally, in the analog domain, $\textit{I\textsubscript{AVDD}}$ can be increased until the thermal noise level is reduced to a satisfactory level. $\textit{V\textsubscript{AVDD}}$ is chosen considering the device operating point and required dynamic range. In the digital domain, $\textit{I\textsubscript{DVDD}}$ is determined by a clock speed (or operating frequency). $\textit{V\textsubscript{DVDD}}$ can be set relatively lower than $\textit{V\textsubscript{AVDD}}$ and reduced to a level that guarantees reliable operations of the logic gates. Revisiting Eq. (40), $V\rlap{\textsuperscript{2}}\textit{\textsubscript{n,Dout}}$ can be rearranged as follows

\begin{align}
	V\rlap{\textsuperscript{2}}\textit{\textsubscript{n,Dout}} &=
	\big(V\rlap{\textsuperscript{2}}\textit{\textsubscript{n,analog}} + V\rlap{\textsuperscript{2}}\textit{\textsubscript{n,DAC}}\big)\frac{G\rlap{\textsuperscript{2}}\textit{\textsubscript{analog}}}{\big(1 + \textit{G\textsubscript{analog}}\big)^2} + V\rlap{\textsuperscript{2}}\textit{\textsubscript{n,qtz}}\frac{1}{\big(1 + \textit{G\textsubscript{analog}}\big)^2}
\end{align}

\noindent
Assuming that $\textit{G\textsubscript{analog}}$ is high enough in Eq. (45), the second term can be negligible and the first term is approximated as $V\rlap{\textsuperscript{2}}\textit{\textsubscript{n,analog}} + V\rlap{\textsuperscript{2}}\textit{\textsubscript{n,DAC}}$. Eventually, assuming $\textit{G\textsubscript{analog}}$ is sufficiently high, the following result is obtained from Eqs. (43) and (45):

\begin{align}
	V\rlap{\textsuperscript{2}}\textit{\textsubscript{n,DCFE,in}} \approx V\rlap{\textsuperscript{2}}\textit{\textsubscript{n,Dout}} \approx \big(V\rlap{\textsuperscript{2}}\textit{\textsubscript{n,analog}} + V\rlap{\textsuperscript{2}}\textit{\textsubscript{n,DAC}}\big)
\end{align}

\begin{figure}[h]
	\centering
	\includegraphics[scale=0.6]{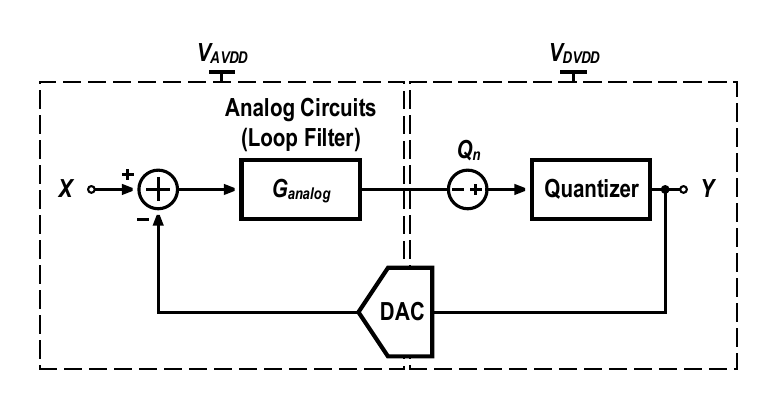}
	\caption{Basic operation of a DCFE including a noise source.}
	\label{Fig. 7}
\end{figure}

\indent
This result can be also explained through the basic operation of a simple delta-sigma ($\Delta\Sigma$) modulator. To obtain the transfer functions, Fig. \ref{Fig. 7} redraws Fig. \ref{Fig. 6}(c) including the noise source $\textit{Q\textsubscript{n}}$, input signal $X$, and output signal $Y$. $\textit{Q\textsubscript{n}}$ refers to the overall noise voltage appearing at the input node of the quantizer, which is the sum taking into account all noise sources of the loop filter, quantizer, and DAC. Note that Fig. \ref{Fig. 7} illustrates a conceptually simplified diagram of DCFE operating in the voltage domain, although an actual DCFE is designed complexly using the elaborately designed blocks such as the loop filter, DAC, and quantizer to achieve high performance. However, this simplified diagram is helpful to understand the noise properties at the input and output nodes of a DCFE.

\indent
In the feedback path in Fig. \ref{Fig. 7}, the DAC copies $Y$ accurately to the negative feedback node. Then, $X - Y$ is generated at the input node of the loop filter. The loop filter is interchangeably used with the analog circuits as shown in Fig. \ref{Fig. 7}. Combining the signal and noise source, the output signal $Y$ can be generated and organized as follows:

\begin{align}
	Y &= \textit{G\textsubscript{analog}}\big(X - Y\big) + \textit{Q\textsubscript{n}}\\
	Y\big(1 + \textit{G\textsubscript{analog}}\big) &= \textit{G\textsubscript{analog}}X + \textit{Q\textsubscript{n}}\\
	Y &= \frac{\textit{G\textsubscript{analog}}}{1 + \textit{G\textsubscript{analog}}}X + \frac{1}{1 + \textit{G\textsubscript{analog}}}\textit{Q\textsubscript{n}}
\end{align}

\noindent
where $\textit{G\textsubscript{analog}}/(1 + \textit{G\textsubscript{analog}})$ is the signal transfer function and $1/(1 + \textit{G\textsubscript{analog}})$ is the noise transfer function, which are used in Eqs. (40) and (41). These equations are well-known and widely used to describe the operation of a $\Delta\Sigma$ modulator. Interestingly, the output noise form of Eq. (45) is the same as the output signal form of Eq. (49): (1) the sum of noise sources, $V\rlap{\textsuperscript{2}}\textit{\textsubscript{n,analog}} + V\rlap{\textsuperscript{2}}\textit{\textsubscript{n,DAC}}$, at the DCFE input is converted by $\textit{G\textsubscript{analog}}/(1 + \textit{G\textsubscript{analog}})$, and (2) the noise source $V\rlap{\textsuperscript{2}}\textit{\textsubscript{n,qtz}}$ at the quantizer input is converted by $1/(1 + \textit{G\textsubscript{analog}})$.

\indent
As shown in Eq. (49) and Fig. \ref{Fig. 7}, the input signal $X$ is converted to the output signal $Y$ by the gain of $Y/X (= \textit{G\textsubscript{analog}}/(1 + \textit{G\textsubscript{analog}}))$. This means that $X$ is exactly copied to $Y$ if $\textit{G\textsubscript{analog}}$ is sufficiently high. The amplitude of $Y$ is therefore equal to that of $X$. In terms of the noise perspective between the input and output nodes, the amplitude of the output noise is the same as the amplitude of the input noise if $\textit{G\textsubscript{analog}}$ is high enough, resulting in $V\rlap{\textsuperscript{2}}\textit{\textsubscript{n,DCFE,in}} \approx V\rlap{\textsuperscript{2}}\textit{\textsubscript{n,Dout}} \approx \big(V\rlap{\textsuperscript{2}}\textit{\textsubscript{n,analog}} + V\rlap{\textsuperscript{2}}\textit{\textsubscript{n,DAC}}\big)$. Similar to the noise properties of MMRS, the input noise of a DCFE is also dominated by the circuit blocks placed closest to the input signal if the gain of the preceding stage is sufficiently high, e.g., $\textit{G\textsubscript{analog}}$ of DCFE and $G\textsubscript{1}G\textsubscript{2}$ of MMRS. However, in practical front-end design, $\textit{G\textsubscript{analog}}$ is a finite value rather than an infinite value. Therefore, the quantization noise cannot be ignored in the input and output noise voltages shown in Eqs. (43) and (45).

\subsection{\textit{Other Design Parameters}}
Although the current-noise efficiency and power-noise efficiency are only discussed in this article, there are still many other design parameters that need to be considered in the front-end design. For example, (1) a wide dynamic range of the front-end is essential to measure weak voltage signals and artifacts having a large amplitude, (2) a high input impedance of the front-end is required considering a high source impedance from the electrodes, (3) a design area of the single front-end needs to decrease for miniaturization of implantable devices, etc.

\begin{figure}[h]
	\centering
	\includegraphics[scale=0.7]{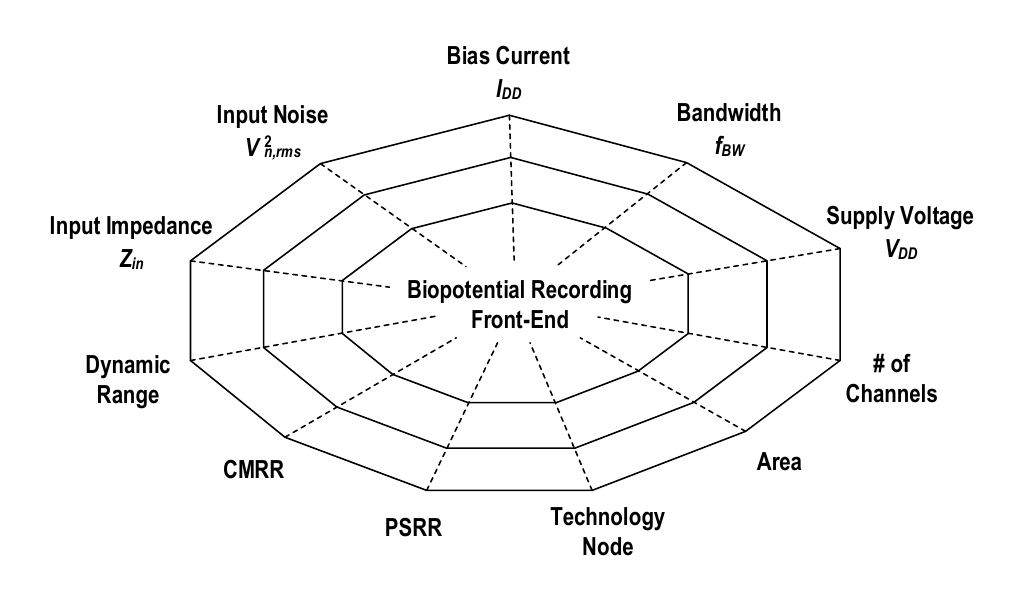}
	\caption{Key design parameters of the biopotential recording front-end.}
	\label{Fig. 8}
\end{figure}

\indent
Fig. \ref{Fig. 8} summarizes the key design parameters that need to be considered in the biopotential recording front-end design. The NEF describing the current-noise efficiency of a front-end is developed using the input noise, bias current, and bandwidth. The PEF is obtained by combining the supply voltage with the parameters used to calculate the NEF. Most of the parameters shown in Fig. \ref{Fig. 8} encounter trade-offs, presenting challenges in the high-performance front-end design.

\indent
For example, a low-noise front-end is essential to detect a weak voltage signal and a certain amount of bias current is required, which is then related to the increase in power consumption. As one way to reduce the power consumption while maintaining low noise, the supply voltage is reduced to a certain level that guarantees the operating point of transistors. However, reducing the supply voltage could degrade the dynamic range. Also, to achieve a low noise while consuming a small current, the chopper can be used. However, the chopper could degrade the input impedance and require additional circuit blocks to suppress signal distortion caused by the chopper. For these reasons, improving the performance of a specific parameter could trade with other performances, and a comprehensive design strategy is required considering target performance and applications.

\section{\Large{C}\large{ONCLUSION}}
In this article, NEF and PEF are discussed from the equation derivation to the trend survey. Both current-noise efficiency and power-noise efficiency are obtained with respect to a single-ended common-emitter stage and a differential common-emitter stage. To provide a comprehensive performance comparison of the front-ends, the performance mapping is developed using the design parameters of the technology node, $\text{NEF}$, $\text{PEF}$, $|\text{PEF} - \text{NEF}|$, and $\textit{V\textsubscript{DD}}$. Using $|\text{PEF} - \text{NEF}|$ provides that how well a front-end balances between current-noise efficiency and power-noise efficiency, in other words, how biased a front-end is between current- and power-noise efficiencies. Also, the performance mapping of each front-end architecture is presented.

\indent
In modern biomedical engineering, especially in the field of biopotential recording, multi-channel multi-stage recording systems have been widely employed to handle a significant amount of bioelectric signals while maximizing system operation efficiency. As semiconductor technology advances and better front-end performance is required, the use of the DCFE has also increased. The noise analysis is therefore conducted for the multi-channel multi-stage recording system and DCFE. Finally, this article briefly addresses the key design parameters to consider in the front-end design, even though this article does not mention in detail beyond NEF and PEF.



\singlespacing

\noindent
\textbf{Taeju Lee} received the B.S. degree in Electrical, Electronics and Communication Engineering from Korea University of Technology and Education (KOREATECH), Cheonan-si, Republic of Korea, in 2014, the M.S. degree in Information and Communication Engineering from Daegu Gyeongbuk Institute of Science and Technology (DGIST), Daegu, Republic of Korea, in 2016, and the Ph.D. degree in Electrical Engineering from Korea Advanced Institute of Science and Technology (KAIST), Daejeon, Republic of Korea, in 2021.

\indent
From 2021 to 2022, he was a Postdoctoral Researcher at KAIST, Daejeon, Republic of Korea. Since 2022, he has been a Postdoctoral Research Scientist at Columbia University, New York, NY, USA. His current research interests include integrated circuits (ICs) used in neuromodulation technology, ultrasound systems, and computing engines.
\end{document}